\documentclass[
prd,
preprint,showpacs,nofootinbib,preprintnumbers]{revtex4}

\usepackage{graphicx}

\usepackage{feynmf}

\usepackage{amssymb}

\usepackage{pifont}

\def\a{\alpha}  \def\b{\beta} \def\g{\gamma} \def\G{\Gamma}

\def\d{\delta} \def\D{\Delta} \def\e{\epsilon}

   \def\m{\mu}

  \def\p{\pi}

 \def\O{\Omega}

\def\bfnabla{\mbox{\boldmath $\nabla$}}

\def\bfsigma{\mbox{\boldmath $\sigma$}}

\def\lQ{\Lambda_{\rm QCD}}

\newcommand{\be}{
\begin{equation}
}{\bf }

\newcommand{\ee}{
\end{equation}
}

\newcommand{\bea}{
\begin{eqnarray}
}

\newcommand{\eea}{
\end{eqnarray}
}

\newcommand{\nn}{\nonumber}

\def\als{\alpha_{\rm s}}

\def\siml{{\ \lower-1.2pt\vbox{\hbox{\rlap{$<$}\lower6pt\vbox{\hbox{$\sim$}}}}\ }}

\begin{document}

\title{Nonrelativistic bound states at finite temperature (I): \\the hydrogen atom}

\author{Miguel \'Angel Escobedo}

\author{Joan Soto}

\affiliation{Departament d'Estructura i Constituents de la Mat\`eria and Institut de Ci\`encies del Cosmos, Universitat de Barcelona\\
Diagonal 647, E-08028 Barcelona, Catalonia, Spain}

\preprint{UB-ECM-PF 08/05}

\pacs{11.10.St,11.10.Wx,12.20.Ds,31.30.jf,12.38.Mh,14.40.Nd}

\begin{abstract}

We illustrate how to apply modern effective field theory techniques and dimensional regularization to factorize the various scales that appear in nonrelativistic bound states at finite temperature. We focus here on the simplest case: the hydrogen atom. We discuss in detail the interplay of the hard, soft and ultrasoft scales of the nonrelativistic system at zero temperature with the additional scales induced at finite temperature. We also comment on the implications of our results for heavy quarkonium bound states in the quark gluon plasma.

\end{abstract}

\maketitle

\section{Introduction}

Finite-temperature effects in atoms were an issue in the early 1980s \cite{farley,PalanquesMestre:1984xw,Donoghue:1984zz,Cha:1984tz}. The basic physics at low temperatures was already understood in those days \cite{farley} and some experiments displaying finite-temperature effects were successfully carried out \cite{hollberg}. The motivation for reconsidering this topic is that QED bound states are a good testing ground for heavy quarkonium physics \cite{Brambilla:2004wf}. Indeed quite some number of effective theory techniques, including the use of dimensional regularization, were first tested in QED \cite{Caswell:1985ui,Pineda:1997bj,Pineda:1997ie,Pineda:1998kn,Czarnecki:1999mw,Manohar:2000rz,Pineda:2002bv}, and have now become standard tools in heavy quarkonium physics (see \cite{Brambilla:2004jw} for a review). The behavior of heavy quarkonia states at finite temperature has been believed for a long time to be a good probe of the so-called quark gluon plasma \cite{Matsui:1986dk} (see \cite{Lourenco:2006sr} for a recent a overview). With the advent of current experiments at the Relativistic Heavy Ion Collider (RHIC) and the Large Hadron Collider (LHC), precision in the quantification of this phenomenon will be necessary, and hence computational tools must be developed. A number of works in this direction have recently appeared in the literature \cite{Laine:2006ns,Laine:2007gj,Burnier:2007qm,Beraudo:2007ky,na}.

We present in this paper an efficient way to include finite-temperature effects in nonrelativistic bound states. We focus here on the simplest of them, namely, the hydrogen atom, and make extensive use of nonrelativistic QED (NRQED) \cite{Caswell:1985ui} and Potential NRQED (pNRQED) \cite{Pineda:1997bj,Pineda:1997ie,Pineda:1998kn}. Since these effective theories are based on momentum expansions about the on-shell condition, which do not exist in Euclidean space, it is compulsory to use the real time formalism (see, for instance, \cite{lebellac}).

In the hydrogen atom, complications due to hard thermal loops (HTLs) \cite{Frenkel:1989br,Braaten:1990az,Taylor:1990ia,Braaten:1991gm} can be ignored at low temperatures ($T\ll m$, $m$ being the electron mass). This allows one to carry out precision calculations in two relevant regimes, namely, when $T \lesssim E$, $E$ being the binding energy, and $T \sim p >> E$, where $p$ is the typical momentum of the electron ($\sim$ inverse Born radius). We critically compare with previous results in the literature. Then we move to the high-temperature case $T\sim m$, which, to our knowledge, has not been studied before. We carry out the matching from QED to NRQED at finite temperature and discuss the effects of the HTL in the bound state dynamics.

We distribute the paper as follows. In the next section we review the two effective theories mentioned above, which are extremely useful for the description of QED bound states at zero temperature, and discuss how they are affected by a finite temperature. In Sections III, IV and V we address the cases $T\sim E$, $T\sim p$ and $T\sim m$ respectively. Section VI is devoted to a  discussion of our results and to some conclusions. Three Appendixes contain technical details.

\section{The hydrogen atom}

The relevant (energy) scales in the states of principal quantum number $n$ of a hydrogen atom at $T=0$ are the electron mass $m$ (hard), the inverse Born radius $p=m\alpha/n$ (soft) and the binding energy $E=-m\alpha^2/2n^2$ (ultrasoft). They satisfy the inequalities $m \gg p \gg E$, which are most conveniently exploited using effective field theories. NRQED is the effective theory which exploits the inequality $m \gg p , E$. It is obtained from QED by integrating out momentum scales of order $m$ and is equivalent to it at any desired order in the $p/m$, $E/m$ and $\alpha$ expansions. It reads

\begin{eqnarray}
\mathcal{L}=\psi^+(iD^0+\frac{{\bf D}^2}{2m}+\frac{{\bf D}^4}{8m^3}+c_F e\frac{{\bf \sigma}{\bf B}}{2m}+c_D e\frac{|{\bf \nabla}{\bf E}|}{8m^2}+\\
+ic_S e\frac{{\bf \sigma}({\bf D}\times{\bf E}-{\bf E}\times{\bf D})}{8m^2})\psi+N^+iD^0N-\frac{1}{4}F_{\mu\nu}F^{\mu\nu}+\frac{d_2}{m^2}F_{\mu\nu}D^2F^{\mu\nu}\nn
\end{eqnarray}

$D_\mu=\partial_\mu+ieA_\mu$ ($D_\mu=\partial_\mu-iZeA_\mu$, $Z$ is the charge of the nucleus) when acting on $\psi$ ($N$), {\bf E} ({\bf B}) is the electric (magnetic) field, and $c_D$, $c_F$, $c_S$ and $d_2$ are matching coefficients, which encode the nonanalytic dependence on the scale $m$. At ${\cal O}(\a)$ they read,\cite{Manohar}

\begin{equation}
c_D=1+\frac{\alpha}{\pi}\left(\frac{8}{3}\ln\frac{m}{\mu}\right)+...\nn\label{cd}
\end{equation}

\begin{equation}
c_F=1+\frac{\alpha}{2\pi}+...\nn
\end{equation}

\begin{equation}
c_S=1+\frac{\alpha}{\pi}+...
\end{equation}

\begin{equation}
d_2=\frac{\alpha}{60\pi}+...\nn
\end{equation}

The remaining inequality, $p\gg E$ is most conveniently exploited using pNRQED. pNRQED is obtained from NRQED by integrating out energy scales of order $p$ and 
it is equivalent to it at any desired order in $E/p$ and $\alpha$. It reads

\begin{eqnarray}
L_{pNRQED}=\int\,d^3{\bf x}(\psi^+\{iD^0+\frac{{\bf D}^2}{2m}+\frac{{\bf D}^4}{8m^3}\}\psi+N^+iD^0N-\frac{1}{4}F_{\mu\nu}F^{\mu\nu})+\nn\\
+\int\,d^3{\bf x}_1\,d^3{\bf x}_2N^+N(t,{\bf x}_2)(\frac{Z\alpha}{|{\bf x}_1-{\bf x}_2|}+\frac{Ze^2}{m^2}(-\frac{c_D}{8}+4d_2)\delta^3({\bf x}_1-{\bf x}_2)+\label{1pnr}\\
+ic_S\frac{Z\alpha}{4m^2}{\bfsigma}(\frac{{\bf {\bf x}_1}-{\bf {\bf x}_2}}{|{\bf x}_1-{\bf x}_2|^3}\times{\bfnabla}))\psi^+\psi(t,{\bf x}_1)\nn
\end{eqnarray}

The potentials above play the role of matching coefficients, which encode the nonanalytic dependence on the scale $p$.
The photon fields in the covariant derivatives contain only energy and momentum much smaller than $p$. This Lagrangian can be written in a manifestly gauge invariant form in terms of a wave-function field $S (t,{\bf x})$, which describes an ion of charge $(Z-1)e$ and gauge transforms with respect to the center of mass only (it is gauge invariant for $Z=1$;
see \cite{Pineda:1997ie} for details).
It reads

\bea
&L_{pNRQED}=&- \int d^3{\bf x} {1\over 4} F_{\mu \nu} F^{\mu \nu}
+\int d^3{\bf x} S^{\dagger} (t,{\bf x})\Biggl(iD_0
+ {{{\bfnabla}}^2\over 2 m} 
+{Z\a\over \vert {\bf x}\vert}
+
\nn
\\&&
+ {{{\bfnabla}}^4\over
8 m^3}
 +{Ze^2\over m^2}\left( -{c_D\over 8} +4d_2\right)\d^3 ({\bf x}) 
 +ic_S{Z\a\over 4m^2}{\bfsigma} \cdot \left({{\bf x}\over \vert {\bf x} 
\vert^3}\times {\bfnabla} \right)
\Biggr)S (t,{\bf x})
\label{pnrqcd2}
\\ && 
+\int d^3{\bf x} S^{\dagger} (t,{\bf x})e{\bf x} \cdot {\bf E}S (t,{\bf x})\nn
\,.
\eea
The size of each term above can be obtained using ${\bfnabla}\sim \vert {\bf x}\vert^{-1}\sim m\a$, $i\partial_0\sim m\a^2$ and ${\bf E}\sim m^2\a^4$ ($Z\sim 1$ will be assumed for the estimates throughout). The leading order terms are then in the first line, and produce the well known Coulomb spectrum at ${\cal O}(m\a^2)$.
The spectrum at $O(m\alpha^5)$ can easily be calculated from the Lagrangian above, by treating the remaining terms as perturbations. The calculation is divided into two parts: (i) a standard quantum mechanical calculation of the expectation value of the potentials in the middle line between the Coulomb states and (ii) the contribution of the ultrasoft (US) photons, which arise from perturbations involving the last term. The former gives,
\bea
\d^{S} E_{n} =&&\d^{S,K} E_{n} +\d^{S,\d} E_{n} +\d^{S,S} E_{n}\nn \,,\\&&\label{s0}\\
\d^{S,K} E_{n} =&&-{1 \over 8m^3} \langle nlj\vert \bfnabla^4 \vert nlj \rangle \nn \,,\\
\d^{S,\d} E_{n} =&&{Ze^2\over m^2} \left( {c_D\over 8}-4d_2 \right) \vert
\phi_{n}({\bf 0})\vert^2 \nn \,,\\
 \d^{S,S} E_{n}  =&&c_S{Z\a\over 4m^2} \left( j(j+1)-l(l+1)-{3\over 4}\right)
\langle nlj\vert {1\over{\bf x}^3}\vert nlj \rangle 
\nn 
\eea
($\vert n \rangle =\vert nlj \rangle $), and the latter
\bea
\d^{US} E_{n} = && \frac{4Z\alpha^2}{3}\Bigg(\left( \ln{\m\over m}+{5\over
6}-\ln{2}\right){\vert \phi_{n}({\bf 0})\vert^2 \over m^2}
\nn\\
 && \label{us0}\\
&&
-\sum_{r\not= n}
\vert
\langle n\vert {\bf v}\vert r \rangle
\vert^2
\left( E_{n} -E_{r}\right)\ln{m\over\vert  E_{n} -E_{r}\vert} \Bigg)
\nn
\,.
\eea
together with the total width
\be
\G_n=\sum_{r<n}{4\over 3}\a
\vert
\langle n\vert {\bf v}\vert r \rangle 
\vert^2 \left(E_n -E_r \right)
\,.\label{width0}
\ee
where ${\bf v}=-i{\bfnabla}/m$ and $\phi_{n}({\bf 0})$ is the wave function at the origin. The correction to the total energy is given by
\be
\d E_n =\d^{S} E_{n} + \d^{US} E_{n}
\,,\label{e0}
\ee
in which the $\mu$ dependence is canceled between the ultrasoft contribution and the one in $c_D$; see (\ref{cd}).

At finite $T$, we have to find out how to properly account for the new scale $T$. The first important property, which follows from the Boltzmann distribution, is that fluctuations of energy much larger than $ T$ are exponentially suppressed. This implies that for $m \gg T$ the same NRQED Lagrangian as for $T=0$ can be used: the temperature dependence of the hard matching coefficients is exponentially suppressed and hence negligible. It also implies that for $p \gg T$ the same pNRQED Lagrangian as for $T=0$ can be used: the temperature dependence of the potentials is exponentially suppressed and hence negligible. We begin by analyzing this case, in which finite-temperature effects are encoded in the ultrasoft photons, in the following section. Next we move on to the case $m\gg T \gg E$. In this case the finite-temperature effects must be taken into account in the matching between NRQED and pNRQED, and are encoded in temperature-dependent potentials. For $T \sim m$, the finite-temperature effects must already be taken into account in the matching between QED and NRQED, and are encoded in the temperature-dependent NRQED matching coefficients and in the HTL effective Lagrangian. As in the $T=0$ case, we will use the Coulomb gauge for calculations in NRQED and pNRQED, and the Feynman gauge for calculations in QED.

\section{The case $p \gg T$}\label{Tus}

As mentioned before, we can just consider the pNRQED Lagrangian at zero temperature. The finite-temperature effects are encoded in the ultrasoft photons, and not in the potentials, which remain the same as in the zero-temperature case. Let us count $T \sim E$ and present the calculation at order $m\alpha^5$. If we use the Lagrangian (\ref{1pnr}), there are two contributions to the binding energy (and decay width). The first one is given by the photon tadpole arising from the kinetic term. It reads ($\beta=1/T$),

\be
\parbox{30mm}{\begin{fmffile}{./tmaster5}
\begin{fmfgraph}(80,80)
\fmfleft{i}
\fmfright{o}
\fmf{fermion}{i,v}
\fmf{photon,right=90}{v,v}
\fmf{fermion}{v,o}
\end{fmfgraph}
\end{fmffile}}
=-\frac{i\pi\a}{3m\beta^2}\label{tadpole}.
\ee
The wavy line stands for the tranverse photon propagator (in the Coulomb gauge), and the solid line for the nonrelativistic electron propagator. This contribution
 is bound state independent and coincides with the thermal mass shift obtained in direct QED calculations \cite{Donoghue:1984zz}. The second
contribution is given by calculating the following ultrasoft loop
at finite temperature.

\bea
\parbox{30mm}{\begin{fmffile}{./tmaster22}
\begin{fmfgraph*}(80,80)
\fmfleft{i}
\fmfright{o}
\fmf{heavy,label=p}{i,v1}
\fmf{heavy,label=p-k}{v1,v2}
\fmf{photon,label=k,left,tension=0.1}{v1,v2}
\fmf{heavy,label=p}{v2,o}
\end{fmfgraph*}
\end{fmffile}} & &    =-e^2\lim_{p_0\to E_n}\langle n|v^i I_{ij}(p_0-H_0) v^j|n\rangle
\label{ust}
\eea

\be
=-e^2\lim_{p_0\to E_n}\sum_r\langle n|v^i|r\rangle I_{ij}(p_0-E_r)\langle r|v^j|n\rangle .
\ee

The double line indicates that the Coulomb potential is taken into account exactly in the propagator, and
\be
H_0=-{{{\bfnabla}}^2\over 2 m} 
-{Z\a\over \vert {\bf x}\vert}
\quad ,\quad\quad
  I_{ij}(q)=\int\frac{\,d^4k}{(2\pi)^3}\frac{\delta(k^2)}{e^{\beta|k_0|}-1}(\delta_{ij}-\frac{k_ik_j}{{\bf k}^2})\frac{i}{q-k_0+i\eta}.
\label{usI}
\ee

We have displayed only the temperature-dependent piece [the temperature-independent one has already been given in (\ref{us0}) and (\ref{width0})]. If the gauge invariant formulation (\ref{pnrqcd2}) is used instead, the whole contribution comes from the last ultrasoft loop. Separating (\ref{usI}) into real and imaginary parts, we obtain,

\be
\Re I_{ij}(q)=\frac{\delta_{ij}}{6\pi}\frac{|q|}{e^{\beta|q|}-1} , \label{usre}
\label{re}
\ee

\be
\Im I_{ij}(q)=\frac{\delta_{ij}q}{6\pi^2}(\ln(\frac{2\pi}{\beta|q|})+\Re\psi(\frac{i\beta|q|}{2\pi})) . \label{usim}
\label{im}
\ee
The intermediate calculations for the imaginary part are presented in Appendix A. We have not been able to proceed analytically any further in the general case. We may write down our final results for the thermal energy shift and decay width in terms of (\ref{re}) and (\ref{im}) as

\bea
\d E_n&=&\frac{\pi\a}{3m\beta^2}+e^2\sum_r\vert \langle n|{\bf v}|r\rangle\vert^2 \Im I_{ij}(E_n-E_r) , \label{e}\\
\d \G_n&=&2e^2\sum_r\vert \langle n|{\bf v}|r\rangle\vert^2 \Re I_{ij}(E_n-E_r) . \label{g}
\eea
These final expressions are suitable for numerical treatment. 
Further analytical results can be obtained in the limiting cases $E\gg T$ and $E\ll T$
, which we present below.

\subsection{$E\gg T$}

In this case, the real part (\ref{usre}) is exponentially suppressed, and hence no temperature-dependent contribution to the decay width (\ref{g}) arises. The imaginary part
can be obtained by expanding $\Re\psi(iy)$ for large $y$ in (\ref{usim}),

\be
\Re\psi(iy)=\Re\psi(1+iy)\sim \ln(y)+\frac{1}{12y^2}+\frac{1}{120y^4}+...
\label{lowT}
\ee
or alternatively $k$ over $q=E_n-H_0$ in the integrand of (\ref{usI}). The leading contribution reads

\be
I_{ij}=\frac{i\delta_{ij}}{18\beta^2q},
\ee
so

\be
\delta E_n=-\frac{2}{9}\frac{\pi\alpha}{\beta^2}\langle n|{\bf v}\frac{\bar P_n}{H_0-E_n}{\bf v}|n\rangle=-\frac{\pi\alpha}{3m\beta^2},
\ee
$\bar P_n=1-P_n$, $P_n$ is the projector onto the subspace of energy $E_n$ (note that $I_{ij}(0)=0$). This contribution cancels exactly that of the photon tadpole (\ref{tadpole}), namely, the first term in the right-hand side of (\ref{e}).  This cancellation appears to be automatic if one uses the gauge invariant Lagrangian (\ref{pnrqcd2}). Either way, the leading nonvanishing contribution comes from the third term in (\ref{lowT}),

\be
\delta E_n=-\frac{4\pi^3\alpha}{45\beta^4}\langle n|{\bf v}\frac{\bar P_n}{(H_0-E_n)^3}{\bf v}|n\rangle .
\label{vl0}
\ee
The matrix element above can be evaluated analytically using the techniques of \cite{Voloshin:1979uv,Pineda:1998id}. We obtain,


\be
\langle n|{\bf v}\frac{\bar P_n}{(H_0-E_n)^3}{\bf v}|n\rangle=\langle n|{\bf x}\frac{\bar P_n}{(H_0-E_n)}{\bf x}|n\rangle=\frac{l}{2l+1}A(n,l)+\frac{l+1}{2l+1}B(n,l) ,
\label{vl}
\ee
where

\be
A(n,l)=\frac{1}{64nmE_n^2}\{F(n,-l-1)-F(-n,-l-1)+2(n^2-l^2)[-24(5n^2-l^2+1)+3n(20l+15)]\},
\ee
and
\be
B(n,l)=\frac{1}{64nmE_n^2}\{F(n,l)-F(-n,l)+2(n^2-(l+1)^2)[-24(5n^2-(l+1)^2+1)+3n(28l+67)]\},
\ee
with

\be
F(n,l)=(n+l+2)(n+l+1)[\frac{(n+l+3)(n-l)}{2}+4(2n-l)^2].
\ee
The details of this computation are given in Appendix (A.2).
Note the strong dependence of the expression above on the principal quantum number $\sim {n^6/ m^3\a^4}$.
Let us then summarize our final results for the thermal energy shift and decay width in this case as

\bea
\d E_n&=& -\frac{4\pi^3\a}{45\beta^4}\langle n|{\bf x}\frac{\bar P_n}{(H_0-E_n)}{\bf x}|n\rangle\label{ets}\left(1+{\cal O}\left(({n^2\over \b m\a})^2\right)\right) , \\
\d \G_n&=& 0 . \label{gts}
\eea

\subsection{$E\ll T$}

In this case, the real part can be easily evaluated by expanding the exponential. At leading order in this expansion, it leads to an additional temperature-dependent decay width for all the states.

\be
\delta \Gamma_n={4Z^2\a^3\over 3\b n^2} . \quad 
\label{widtht}
\ee
The total width is obtained by summing the $T=0$ contribution (\ref{width0}) to the expression above.
The imaginary part
is obtained by doing the $y\rightarrow 0$ expansion in (\ref{usim}),

\be
\Re \psi (iy)=-\gamma +{\cal O} (y^2) .
\ee
Alternatively, one may expand $q=E_n-H_0$ over $k$ in (\ref{usI}). Then 
the Bethe logarithms from (\ref{usim}) cancel out against those of the zero temperature contribution (\ref{us0}),
and we get for 
the whole ultrasoft contribution,

\be
\delta^{US} E_n=\frac{4Z\alpha^2}{3}(\ln(\frac{\beta\mu}{2\pi})+\frac{5}{6}-\ln 2+\gamma)\frac{|\phi_n(0)|^2}{m^2} .
\label{ustus}
\ee 
The total binding energy is obtained from (\ref{e0}) using the expression above for $\delta^{US} E_n$ and (\ref{s0}) for $\delta^{S} E_n$.
Alternatively, we may summarize our final results for the thermal energy shift and decay width in this case as

\bea
\d E_n &=& \frac{\alpha\pi}{3m\beta^2}+\frac{2\alpha}{3\pi}\sum_r\vert \langle n|{\bf v}|r\rangle\vert^2 
(E_n-E_r)( \ln(\frac{2\pi}{\beta|E_n-E_r|})-\g )
(1+{\cal O}
\left(({\b m\a\over n^2})^2\right)
) , \label{ets}\\
\d \G_n &=& {4Z^2\a^3\over 3\b n^2}\left(1+{\cal O}\left({ \b m\a\over n^2}\right)\right)   \label{gtl} . 
\eea

\section{$m\gg T \gg E$}

In this case finite-temperature effects are expected to modify the potential, which might in principle give rise to qualitatively different effects. However, for QED at energies below the electron mass the vacuum polarization effects are suppressed by even powers of $m$, and hence the full $A_0$ propagator in the Coulomb gauge is not sensitive to the temperature (up to high orders in $T/m$ ($\sim T^4/m^4$)). Finite-temperature effects enter only through the tranverse photon propagators. Since the coupling of these photons to nonrelativistic electrons is suppressed by powers of $1/m$, the finite-temperature effects modify only the $1/m$ corrections and, hence, the Coulomb potential remains as the leading order term. This implies that the gross features of the hydrogen atom spectrum will be kept the same for temperatures smaller than the electron mass.
We proceed then to the matching between NRQED and pNRQED at finite $T$. At $T=0$ the matching is trivial in the electron sector, since this sector is insensitive to the momentum transfer (to transfer momentum one needs the nucleus), the soft scale to be integrated out. At $T\not=0$ the temperature is a scale to be integrated out and the matching becomes nontrivial in this sector. If we count $T\sim p$, for a calculation at order $m\alpha^5$ we need the contributions of the following diagrams:

\be
\parbox{30mm}{\begin{fmffile}{./tmaster2}
\begin{fmfgraph*}(80,80)
\fmfkeep{self}
\fmfleft{i}
\fmfright{o}
\fmf{fermion,label=p}{i,v1}
\fmf{fermion,label=p-k}{v1,v2}
\fmf{photon,label=k,left,tension=0.1}{v1,v2}
\fmf{fermion,label=p}{v2,o}
\end{fmfgraph*}
\end{fmffile}}
=i\frac{2\alpha p^2}{3\pi m^2}[\ln\frac{\beta\mu}{2\pi}+\gamma-\ln 2+\frac{5}{6}](p_0-\frac{p^2}{2m})-\frac{i\pi\alpha p^2}{9m^3\beta^2},
\ee

\be
\parbox{30mm}{\begin{fmffile}{./tmaster48}
\begin{fmfgraph*}(80,80)
\fmfleft{i}
\fmfright{o}
\fmf{fermion}{i,v1}
\fmfv{label=$c_F$,label.angle=125}{v1}
\fmf{fermion}{v1,v2}
\fmf{photon,left,tension=0.1}{v1,v2}
\fmfv{label=$c_F$,label.angle=60}{v2}
\fmf{fermion}{v2,o}
\end{fmfgraph*}
\end{fmffile}}
=\frac{i\pi\alpha c_F^2}{6m^2\beta^2}(p_0-\frac{p^2}{2m})-\frac{i\alpha\pi^3c_F^2}{30m^3\beta^4},
\ee

\be
\parbox{30mm}{\begin{fmffile}{./tmaster4}
\begin{fmfgraph*}(80,80)
\fmfleft{i}
\fmfright{o}
\fmf{fermion}{i,v1}
\fmfv{label=$c_F$,label.angle=125}{v1}
\fmf{fermion}{v1,v2}
\fmf{photon,left,tension=0.1}{v1,v2}
\fmfv{label=$c_S$,label.angle=60}{v2}
\fmf{fermion}{v2,o}
\end{fmfgraph*}
\end{fmffile}}
=\frac{i\alpha\pi^3c_Fc_S}{60m^3\beta^4},
\ee

\be
\parbox{30mm}{\begin{fmffile}{./tmaster5}
\begin{fmfgraph}(80,80)
\fmfleft{i}
\fmfright{o}
\fmf{fermion}{i,v}
\fmf{photon,right=90}{v,v}
\fmf{fermion}{v,o}
\end{fmfgraph}
\end{fmffile}}
=-\frac{i\pi\alpha}{3m\beta^2},
\ee

\be
\parbox{30mm}{\begin{fmffile}{./tmaster6}
\begin{fmfgraph*}(80,80)
\fmfleft{i}
\fmfright{o}
\fmf{fermion}{i,v}
\fmfv{label=relativistic,label.angle=-90}{v}
\fmf{photon,right=90}{v,v}
\fmf{fermion}{v,o}
\end{fmfgraph*}
\end{fmffile}}
=\frac{i5\alpha\pi p^2}{18m^3\beta^2}.
\ee
The last diagram comes from the $\psi^\dagger {\bf D}^4\psi / {8m^3}$ term in the Lagrangian, which contains a piece with two derivatives and two ${\bf A}$ fields. Other possible diagrams either are of higher order or 
give zero.

The first diagram has an infrared (IR) divergence. We have followed the same prescriptions as in the $T=0$ case. We have regulated it in dimensional regularization (DR) and used the modified minimal subtraction scheme ($\overline{MS}$). When one will eventually make calculations in pNRQED one must regulate the ultraviolet (UV) divergences which will appear there in DR and use the same subtraction scheme. The subtraction point dependence will then cancel out in all observables and the finite pieces will be consistently calculated (see \cite{Pineda:1997ie} for detailed discussions in the $T=0$ case).

The matching in the electron-nucleus sector (i.e., the calculation of the potentials) reduces to the calculation of the following vertex diagrams:

\be
\parbox{30mm}{\begin{fmffile}{./tmaster8}
\begin{fmfgraph}(80,80)
\fmfkeep{vertex}
\fmfleft{i}
\fmfright{o}
\fmfbottom{b}
\fmf{fermion}{i,v1}
\fmf{fermion}{v1,v2}
\fmf{dashes}{v2,b}
\fmf{fermion}{v2,v3}
\fmf{photon,left,tension=0.1}{v1,v3}
\fmf{fermion}{v3,o}
\end{fmfgraph}
\end{fmffile}}
=\frac{2\alpha A({\bf p}{\bf p'})}{3\pi m^2}[\ln\frac{\beta\mu}{2\pi}+\gamma-\ln 2+\frac{5}{6}],
\ee
where
\be
A=
\parbox{30mm}{\begin{fmffile}{./tmaster9}
\begin{fmfgraph}(80,80)
\fmfkeep{tree}
\fmfleft{i}
\fmfright{o}
\fmfbottom{b}
\fmf{fermion}{i,v}
\fmf{dashes}{v,b}
\fmf{fermion}{v,o}
\end{fmfgraph}
\end{fmffile}},
\ee

\be
\parbox{30mm}{\begin{fmffile}{./tmaster11}
\begin{fmfgraph*}(80,80)
\fmfleft{i}
\fmfright{o}
\fmfbottom{b}
\fmf{fermion}{i,v1}
\fmfv{label=$c_F$}{v1}
\fmf{fermion}{v1,v2}
\fmf{dashes}{v2,b}
\fmf{fermion}{v2,v3}
\fmfv{label=$c_F$}{v3}
\fmf{photon,left,tension=0.1}{v1,v3}
\fmf{fermion}{v3,o}
\end{fmfgraph*}
\end{fmffile}}
=\frac{\alpha c_F^2\pi A}{6m^2\beta^2}.
\ee
The dashed line stands for the $A_0$ photon propagator (in the Coulomb gauge).
As before, other possible diagrams are zero or of higher order. Putting all these together we obtain the following temperature dependent corrections to the pNRQED Lagrangian:

\begin{eqnarray}
\delta L(T)=\int\,d^3{\bf x}[\frac{2\alpha}{3\pi m^2}[\ln(\beta\mu)-\ln 2+\frac{5}{6}-\ln(2\pi)-\gamma][\frac{\Delta\psi^+\Delta\psi}{2m}+\Delta\psi^+\partial_0\psi]-\nn\\
-\frac{\pi\alpha}{6m^3\beta^2}{\bfnabla}\psi^+{\bfnabla}\psi+\frac{2\pi\alpha c_F^2}{12m^2\beta^2}[\frac{{\bfnabla}\psi^+{\bfnabla}\psi}{2m}+i\psi^+\partial_0\psi]+(\frac{\alpha\pi}{3m\beta^2})\psi^+\psi]+\nn\\
+\int\,d^3{\bf x}_1\,d^3{\bf x}_2N^+N(t,{\bf x}_2)(\frac{Z\alpha^2c_F^2\pi Z}{6m^2\beta^2|{\bf x}_1-{\bf x}_2|}\psi^+\psi( t,{\bf x}_1)+\\
+\frac{2}{3}\frac{\alpha}{\pi m^2}\frac{Z\alpha}{|{\bf x}_1-{\bf x}_2|}{\bfnabla}\psi^+{\bfnabla}\psi (t,{\bf x}_1)[\ln(\beta\mu)-\ln 2+\frac{5}{6}-\ln(2\pi)-\gamma])\nn ,
\end{eqnarray}
which can be cast into a much simpler form by using the following field redefinition:

\be
\psi \longrightarrow (1+\frac{2\alpha}{3\pi m^2}[\ln\frac{\beta\mu}{2\pi}+\gamma-\ln 2+\frac{5}{6}]\Delta-\frac{\pi\alpha c_F^2}{6m^2\beta^2})\psi ,
\ee

\begin{eqnarray}
\delta L(T)=\int\,d^3{\bf x}[-\frac{\pi\alpha}{6m^3\beta^2}{\bfnabla}\psi^+{\bfnabla}\psi+(\frac{\alpha\pi}{3m\beta^2})\psi^+\psi]+\nn\\
+\int\,d^3{\bf x}_1\,d^3{\bf x}_2 N^+N(t,{\bf x}_2)(-\frac{4Z\alpha^2}{3m^2}(\ln(\frac{\beta\mu}{2\pi})+\gamma-\ln 2+\frac{5}{6})\delta^3({\bf x}_1-{\bf x}_2))\psi^+\psi(t,{\bf x}_1) . \label{LNRT}
\end{eqnarray}
In order to calculate the spectrum at the desired order we only have to sandwich the potentials between the states and calculate the US contribution (and, of course, take into account the relevant mass shifts in (\ref{LNRT})). The first contribution gives

\be
\delta^S E_n=\frac{\alpha\pi}{3m\beta^2}-\frac{\pi\alpha^3}{6m\beta^2 n^2}+\frac{4Z\alpha^2}{3m^2}(\ln(\frac{\beta\mu}{2\pi})+\gamma-\ln 2+\frac{5}{6})|\phi_n(0)|^2 . \label{ests}
\ee
The US contribution corresponds exactly to the diagram (\ref{ust}) , but it has to be calculated taking into account that it contains now only energies much smaller than $T$. In this case the Boltzmann factor can be expanded. This may (and will) introduce UV divergences, which as mentioned before, must be regulated in DR and $\overline{MS}$ subtracted in order to be consistent with the calculation of the potential. We obtain

\be
\Im I_{ij}(q)=q\frac{1}{6\pi^2}\delta_{ij}(
\ln\frac{\mu}{|q|}+\frac{5}{6}-\ln 2)
+{\cal O} (q^3\b^2 ),
\ee
$\Im I_{ij}(q)$ gives a contribution to the binding energy which exactly cancels that of the $T=0$ piece (\ref{us0}).
Then the total binding energy is obtained by adding to (\ref{ests}) the $T=0$ soft contribution (\ref{s0}). $\Re I_{ij}(q)$ gives a contribution to the decay width which coincides with (\ref{widtht}) at leading order in the $m\a /T$ expansion. This contribution is parametrically larger than the zero-temperature decay width (\ref{width0}).
Notice also that in the limit $p\gg T$ the binding energy (\ref{ests}) reduces to (\ref{ustus}), as it should.
We may summarize our final results for the thermal energy shift and decay width in this case as

\bea
\d E_n&=& \frac{\alpha\pi}{3m\beta^2}-\frac{\pi\alpha^3}{6m\beta^2 n^2}+\frac{2\alpha}{3\pi}\sum_r\vert \langle n|{\bf v}|r\rangle\vert^2 (E_n-E_r)(\ln(\frac{2\pi}{\beta|E_n-E_r|})-\g),
\label{es}\\
\d \G_n&=& {4Z^2\a^3\over 3\b n^2}+\frac{2\alpha}{3}\sum_r\vert \langle n|{\bf v}|r\rangle\vert^2  \vert E_n-E_r\vert .
\label{gs}
\eea
The results above are accurate up to corrections of order $m\a^6$ for temperatures $T\sim m\a /n$.

\section{The case $m\sim T$}

For temperatures of the order of the electron mass, electron-positron pairs are created in the thermal bath, which are expected to destabilize the hydrogen atom. In order to make this expectation quantitative, we will integrate out the scale $m \sim T$. In the photon sector, this will induce a mass dependent HTL effective Lagrangian. In the electron sector, not only will the NRQED matching coefficients now depend on $T$, but also new nonlocal terms appear. Let us analyze these two sectors in the following.

\subsection{Matching QED to NRQED+HTL}

\subsubsection{The photon sector}

The HTL effective Lagrangian will be obtained from the vacuum polarization, by standard techniques \cite{lebellac}. Rather than depending on the single scale $eT$, as in the massless case, the HTL effective Lagrangian is now expected to have a nontrivial dependence on $m\b$. In fact, this brings in a new qualitative feature: the angular integration appearing in the massless case becomes a full three-parameter integration \cite{Pisarski:1997cp}. In order to illustrate it, let us focus on the longitudinal component of the retarded self-energy, which will be the only one needed later on. Using the fact that $p_0$, ${\bf p}\ll m$, $T$ and expanding them accordingly we arrive at

\be
\Pi^{00}_R(p)=(-i)2e^2\int {d^3{\bf k}\over (2\pi)^2\sqrt{{\bf k}^2+m^2}}{1\over e^{\b\sqrt{{\bf k}^2+m^2}}+1}{{\bf p}^2-{({\bf p k})^2\over {\bf k}^2+m^2}\over \left( p_0-{{\bf p k}\over \sqrt{{\bf k}^2+m^2}}+i\eta\right)^2}.
\label{rse}
\ee
Note that when $m=0$ the integral over $k=\vert {\bf k}\vert$ decouples from the angular integration and can be carried out analytically. For $m\not=0$, however, the integral over $k$ remains in the effective theory. If we write it in terms of $w:=k/\sqrt{k^2+m^2}$ ($w\in [0,1)$), it is clear that the HTL effective Lagrangian for the photons can be obtained from the one in the massless case (see, for instance, \cite{lebellac}) by doing the following substitutions:

\bea
\hat {\bf k}&\longrightarrow& {\bf w} \\ \nonumber
d\Omega &\longrightarrow& d^3 {\bf w} \\ \nonumber
{\pi^2\over 6\b^2} &\longrightarrow& {2m^2w^2\over (1-w^2)^2\left(e^{\b m\over \sqrt{1-w^2}}+1\right)}
\eea
$\hat {\bf k}={\bf k}/k$ and $d\Omega$ is the integration measure of the solid angle.

\subsubsection{The electron sector (NRQED)}

We have just seen that the photon sector at finite temperature is qualitatively different from the zero-temperature one. Indeed, in the former case a nonlocal HTL effective Lagrangian is produced, which is much more important than the $1/m^2$ suppressed terms that arise at zero temperature (last term in (1)). The question is then whether in the electron sector something similar will also happen. In order to find out, we match QED to NRQED in this sector as follows. We calculate the two-point Green function of an electron with momentum $p_\mu$ and sandwich it between $P_+=\frac{1+\gamma_0}{2}$. Then we make the change $p_0=m+k_0$, ${\bf p}={\bf k}$ and expand for $k_0-\frac{k^2}{2m}$ and ${\bf k}$ small. We will find that, unlike the photon sector, the expansion is local. Then it will be possible to identify $\delta Z_\psi (k)$, the matching coefficient of the nonrelativistic field ($P_+\Psi=\sqrt{ Z_\psi (0)}\psi +{\cal O}(1/m)$, where $\Psi$ stands for the relativistic Dirac spinor field of the electron), and $\Theta (k)$, the NRQED self-energy.

\be
\frac{1+\gamma_0}{2}\parbox{30mm}{\fmfreuse{self}}\frac{1+\gamma_0}{2}=\frac{i\delta Z_\psi (k)}{k_0-k^2/2m}+\frac{i\Theta (k)}{(k_0-k^2/2m)^2}+\cdots
\ee

In the real time formalism, the propagators consist of a sum of the zero temperature part and the thermal part, which will be proportional to $n_B$ for photons and $n_F$ for electrons ($n_{B(F)}$ are the Bose (Fermi) Boltzmann distributions, $n_{B(F)}=1/(e^{\b \vert k_0\vert}\mp 1)$). If we consider just the contribution of $n_B$ ($n_F$), we are taking into account the thermal fluctuations of the photons (electrons). It is important to note that in the diagrams we will consider it will never appear $n_Bn_F$ terms because of kinematic constraints (we will never have an internal electron on shell and an internal photon on shell). Hence we can write $\delta Z_\psi (k)=\delta Z_\psi^B (k)+\delta Z_\psi^F (k)$ and $\Theta (k)=\Theta^B (k)+ \Theta^F (k)$.

Let us first consider the contributions from the thermal fluctuations of the photon to the electron self-energy. 
We obtain

\bea
\Theta^B (k)&=&\frac{\pi\alpha}{3m\beta^2}-\frac{\pi\alpha k^2}{6m^3\beta^2}+{\cal O}\left({k^4\over m^3}\right),
\nn \\
\delta Z_\psi^B (k)&=&{2\a\over \p}\left(I_A+{k^2\over 6m^2}\right)-\frac{\pi\alpha}{3m\beta^2}+{\cal O}\left({k^4\over m^4}\right).
\eea
Note that $\Theta^B (k)$ corresponds to a thermal mass shift $\delta^B m=\pi\alpha/3m\beta^2$ for the electron. $\delta Z_\psi^B$ is IR divergent because of $I_A$:

\be
I_A=\frac{\Omega_{3+\e}}{\Omega_3}\frac{\mu^{-\e}}{(2\pi)^{\e}}\int_0^\infty\frac{\,dq}{q^{1-\e}(e^{\beta q}-1)} =\frac{1}{2}(\gamma+\ln\frac{\beta \mu}{2\pi}-\ln 2),
\ee 
$\O_{D-1}$ is the solid angle in $D-1$ space dimensions, $D=4+\e$, $\e\to 0$, and we have used the $\overline{MS}$ subtraction scheme. $\delta Z_\psi^B$ will be relevant for the calculation presented in Appendix C.

For the thermal fluctuations of the electrons we find a similar result. $\Theta^F (k)$ gives rise to the following thermal mass shift

\be
\delta^F m={4\a m\over \pi} h(m\b )
-{2\a g(m\b) \over\pi m \beta^2},
\label{dm}
\ee
$h(m\b)$ and $g(m\b)$ are defined in Appendix B. Note that $\delta^F m$ above goes to zero exponentially if $m\gg T$. 
$\delta Z_\psi^F (k)$ is simply related to the thermal mass shift $\delta Z_\psi^F (k)=-\delta^F m/m +{\cal O}(k^2/ m^2)$.

In principle we should have taken into account the doubling of degrees of freedoms in this calculation, as we did in the photon sector. However, the off-diagonal components of the self-energy vanish for the same kinematical reasons that forbid terms proportional to $n_B n_F$ above. Hence, the self-energy is diagonal and we can safely ignore the doubling.

In view of the above results, we may wonder if any QED Green function involving electrons will match to local NRQED operators, as is the case of the two-point function, or new nonlocal HTL vertices will arise. Let us then analyze 
the vertex (three-point function with two-electron and one-photon legs) next. The calculations are presented in detail in Appendix C, here we summarize only the more important results.
If we just consider the thermal fluctuations of the photons, 
the vertex can indeed be matched to local NRQED operators. In Appendix C we display the modifications of $c_D$ and $c_S$ in (\ref{cd}) due to temperature in the case $T\ll m$ as an example.
However, if we take into account the thermal fluctuations of the electrons
we get a nonlocal vertex (see (\ref{vnl}) in Appendix C). This vertex is of the same size as the tree level contribution when the momentum transfer $q\sim m\alpha$ (the typical momentum transfer of the bound state at zero temperature), and it is suppressed only by a factor $e$ when $q\sim me$ (the scale of the Debye mass). Hence, it turns out to be much more important than the local contributions arising from the thermal fluctuations of the photons when $T\sim m$. Nevertheless, it goes exponentially to zero when $T\ll m$.

The effective theory for a nonrelativistic electron in a thermal bath of $T\sim m$  (NRQED+HTL) lies then in an intermediate situation between the case $T\ll m$ (NRQED), in which all contributions are local, and the massless case, in which all contributions are nonlocal
(HTL).

\subsection{Matching NRQED to pNRQED with HTLs}

We shall restrict ourselves to the leading order contributions.
The matching is then analogous to the $T=0$ case, which leads to the Coulomb potential, but using the HTL propagator for the ($A_0$) photons. The latter can be obtained from the retarded self-energy ({\ref{rse}) by a standard procedure (see, for instance, \cite{Thoma:2000dc}). It reads

\be
\Delta_{11}(p,p_0)=i(\frac{1}{p^2+m_D^2}-\frac{i16\alpha g(m\beta)}{(p^2+m_D^2)^2p\beta^3}), 
\label{D11}
\ee
where we have used $p_0\ll {\bf p}$, $p=\vert {\bf p}\vert$, and
$m_D^2$
and $g(m\beta)$ are defined in Appendix B.
By Fourier transforming, we obtain the following real space potential:

\be
V=-\frac{Z\a e^{-m_D r}}{r}+\frac{i16Z\a^2g(m\beta)}{\pi m_D^2\beta^3}
\phi(m_D r),
\ee
where
\be
\phi(x)=2\int_0^\infty\frac{\,dz z}{(z^2+1)^2}\left[\frac{\sin(zx)}{zx}\right].
\label{v}
\ee

Unlike for the $T=0$ case, now the $A_0$ photon propagates over arbitrary times, which, together with the fact that its propagator contains scales, implies that contributions to the self-energies of both the electron and the nucleus arise. These read,

\be
\delta m=-\frac{\a m_D}{2}-i\frac{8\a^2g(m\beta)}{\pi m_D^2\beta^3},
\label{ddm}
\ee
for the electron, and the same expression multiplied by $Z^2$ for the nucleus. In order to perform this calculation we need, in principle, $\Delta_{11}(p,p_0)$ for any kinematical region.
However, due to the fact that $\Delta_{11}(p,p_0)=\Delta_{11}(p,-p_0)$ (see, for instance, \cite{lebellac}) we have

\be
\int dp_0 {i\over p_0+i\eta} \Delta_{11}(p,p_0)= \pi \Delta_{11}(p,0) \, ,
\ee
and hence the expression (\ref{D11}) is enough to carry out the calculation.
Formulas (\ref{v}) and (\ref{ddm}) are analogous to the results obtained in \cite{Laine:2006ns} for QCD, which we recover in the $m\rightarrow 0$ limit by setting $Z=1$, and changing $e^2\rightarrow g^2$ and the group factors in $m_D^2$, namely, $1\rightarrow C_A+N_f/2$. Notice, however, that our calculation is much simpler: only one tree-level and one one-loop diagram need to be calculated, instead of the five one-loop diagrams needed in ref. \cite{Laine:2006ns}. It is important to realize that (\ref{v}) has an imaginary part. The HTLs induce the scale $m_D \sim eT$, which for $m\sim T$ dominates over the typical momentum scale of the bound state at $T=0$, $p \sim m\a$, and hence dramatic changes in the bound-state dynamics are expected to occur. Indeed, if $p\sim m_D$ then the imaginary part of the potential is more important than the real one and no bound state is expected to survive. The typical momentum for which the imaginary part becomes of the same order as the real one is $p\sim (16\a)^{1/3}g^{1/3}(m\b ) T=:m_d$. $m_d$ may be considered as a new dynamical scale in the system, which is parametrically larger than $m_D$. Notice that both $m_D$
and 
$m_d$ have a nonanalytic behavior in $m/T$: when $T$ becomes smaller than $m$ they go exponentially to zero. The leading behavior of $m_D$ and $m_d$ for $T < m$ reads

\bea
m_D^2 &\sim & 8 \a \sqrt{\pi m^3\over 2\b }e^{-m\b } , \\
m_d^2 &\sim & \left({16\a m \over \b^2}\right)^{2\over 3}e^{-{2m\b \over 3}} .
\eea
Note that $m_d$ is exponentially larger than $m_D$. This allows us to get more explicit expressions for the energy shift and the decay width of the bound state in the case $m_D \ll m_d\lesssim  p \sim 1/r$, upon expanding (\ref{v}) on $m_D$, and using the asymptotic expressions for $m_D$ and $m_d$ above,

\bea
\delta E_{nl}&=& -{\a m_D (Z-1)^2\over 2}-{Z\a m_D^2\over 2}\langle nl\vert r\vert nl\rangle , \\
\d \G_{nl}&=& 2(Z-1)^2\a\sqrt{2\over \pi \b^3 m} - {2Z\a m_d^3\over 3\pi}\langle nl\vert r^2(\ln m_Dr +\g -{4/3})\vert nl\rangle .
\label{magmD}
\eea
The expressions above hold up to corrections ${\cal O}((m_D n^2/m \a)^2)$. $\langle nl\vert r\vert nl\rangle=[3n^2-l(l+1)]/2Zm\a$ and $\langle nl\vert r^2\vert nl\rangle=[5n^2+1-3l(l+1)]n^2/2(Zm\a )^2$ can be found in standard textbooks, and

\bea 
 \langle nl\vert r^2\ln r \vert nl\rangle &=&{9n\over 2 (Zm\a)^2}{(n-l-1)!\over (n+l)!}\sum_{r=n-l-4}^{n-l-1}{\G (2l+5+r)\over \G (r+1) \G^2 (n-l-r) \G^2 (5+l-n+r)}\times \nn\\
&& \left( \ln {n\over 2Zm\a }+\psi (2l+5+r) + 2\psi (4) - 2\psi (5+l-n+r)\right) .
\eea
which may be obtained using, for instance, the techniques of ref. \cite{Titard:1993nn}.

For $Z\not=1$ it is interesting to observe that the system develops a decay width that is not exponentially suppressed (first term in (\ref{magmD})). This is because a charged ion will tend to capture electrons from the thermal bath to decay into a less charged ion and eventually into a multielectronic atom. Let us focus in the $Z=1$ case. For $n$ large enough, namely, when $m\a /n\sim m_D$, the approximation that leads to (\ref{magmD}) above fails. However, much before, when $m\a /n\sim m_d$, the states $n$ will melt, namely, their decay width will become of the same order as the binding energy. Therefore the expressions in (\ref{magmD}) are appropriated for $T\lesssim m$ as far as it makes sense to speak about states with a narrow width.

\begin{table}[htb]
\makebox[6cm]{\phantom b}
\begin{center}
\begin{tabular}{|c|c|c|c|}
\hline
n & $T_d$ (keV) & $m_D$ (keV) & $m_d$ (keV) \\
\hline
1 & 60.4 & 0.703 & 3.73 \\
2 & 50.1 & 0.284 & 1.86 \\
3 & 45.6 & 0.167 & 1.24 \\
4 & 42.9 & 0.114 & 0.932 \\
5 & 40.9 & 0.0842 & 0.746 \\
\hline
\end{tabular}
\end{center}
\caption{Dissociation temperature, Debye mass $m_D$, and dissociation scale $m_d$,
as a function of the principal quantum number $n$. 
The dissociation (melting) temperature is defined as the temperature for which the dissociation scale $m_d$ equals the soft scale $mZ\a /n$. Note that $m_D$ is smaller than the soft scale but much bigger than the ultrasoft scale $m(Z\a)^2 /n^2$, which is consistent with our assumptions.}
\end{table}

\section{Discussion and Conclusions}

We have developed a formalism which allows us to efficiently factorize the various scales appearing in nonrelativistic bound states at finite temperature. It makes use of dimensional regularization and of the known Effective Field Theories both for nonrelativistic bound states (NRQED,pNRQED) and for finite-temperature systems (HTLs). We have focused on the hydrogen atom.

For $T\ll m$ we have calculated the finite-temperature effects to the binding energy and the decay width to a precision equivalent to ${\cal O}(m\a^5)$. We agree with the early results of \cite{farley}, but disagree with others \cite{PalanquesMestre:1984xw,Donoghue:1984zz,Cha:1984tz}. It is interesting to recall how the finite-temperature effects were experimentally observed in atoms in the early 1980's \cite{hollberg}. Since $E_n\sim m\a^2/n^2$, even if for the ground state $E_1\gg T$, there will always be $n$'s, $n\gg 1$, for which $E_n\ll T$. For the ground state, finite-temperature effects may be very small (given by (\ref{vl0}) and (\ref{vl})) but for highly excited states the thermal mass shift (\ref{tadpole}) must arise. Then transitions from highly excited states to the ground state are sensitive to the thermal mass shift.

For $T\sim m$ we have restricted ourselves to discussing the dominant effects due to finite-temperature. In the photon sector, we have described how to obtain the HTL effective Lagrangian for a finite electron mass. It requires the introduction of an extra integral in addition to the solid angle one. In the electron sector we have seen that in addition to temperature-dependent NRQED matching coefficients, new nonlocal (HTL-like) terms arise. We have calculated the potential at leading order, which develops an imaginary part. The massless limit of this potential agrees with the Abelian limit of the one obtained in \cite{Laine:2006ns}. The imaginary part dominates over the real one for momentum transfer smaller than 
$m_d\sim e^{2/3}Tg^{1/3}(m\beta )$.
For $T < m$, $g(m\beta )$ increases exponentially from zero when $T$ increases. Then, for a given bound state, there will always be a temperature for which the soft scale equals $m_d$, and hence the imaginary part (decay width) equals the real part (energy). We call this temperature dissociation temperature and have calculated it in Table I for the lower-lying states. For temperatures higher than the dissociation temperature, it does not make much sense to speak about a bound state any longer.

We then get the following picture of a hydrogen atom in the ground state when heated from $T=0$ to $T\sim m$. The effects are very small until $T\sim m\alpha^2$. Then it starts developing a width $\sim T\a^3$, which increases with temperature but remains much smaller than the binding energy until $T\sim m$. Then, the width starts increasing exponentially and the hydrogen atom ceases to exist.

From our results we can infer some qualitative features of heavy quarkonium systems in the weak coupling regime (i.e., when the binding is due to a Coulomb-type potential) at finite temperature. These states satisfy $\lQ\lesssim m\als^2$, $\lQ$ being a typical hadronic scale. $\lQ$ affects at most the next-to-leading-order corrections, and hence these states are expected to be rather insensitive to the QCD deconfinement phase transition. When the temperature overcomes the ultrasoft scale ($T > m\als^2$), a decay width proportional to the temperature will be developed, analogously to the hydrogen atom. As the temperature increases further, gluons and light quarks will induce a HTL imaginary part in the potential \cite{Laine:2006ns}, which will become comparable to the real part when $T\sim m \als^{2/3}$. No bound state is expected to survive beyond that temperature. One should keep in mind, however, that only the ground states of bottomonium ($\Upsilon (1S)$ and $\eta_b$), and to a lesser extent of charmonium ($J/\psi$ and $\eta_c$), are likely to be in the weak coupling regime \cite{GarciaiTormo:2005bs}.

\begin{acknowledgments}

We are indebted to Cristina Manuel for introducing us to thermal field theory, for many discussions, encouragement in this project, and the critical reading of the manuscript. We thank Nora Brambilla and Antonio Vairo for making ref. \cite{na} available to us before publication, and for comments on the manuscript. MAE also thanks them for hospitality at the University of Milan while this paper was written up. We acknowledge financial support from the RTN Flavianet MRTN-CT-2006-035482 (EU), the  FPA2007-60275/ and FPA2007-66665-C02-01/ MEC grants (Spain), and the 2005SGR00564 CIRIT grant (Catalonia). MAE has also been supported by MEC FPU (Spain).

\end{acknowledgments}

\appendix

\section{Calculations
in pNRQED}

\subsection{Self-energy}

We proceed to the detailed calculation of the self-energy in pNRQED. It is convenient to separate it into real and imaginary parts. The real part is immediate to obtain and has been given in (\ref{re}), so we will focus on the imaginary part. 
We expand the Boltzmann distribution function in (\ref{usI}) as follows

\be
\frac{1}{e^{\beta k}-1}=-\frac{1}{k}\sum_n\frac{1}{n}\frac{d}{d\beta}e^{-n\beta k},
\ee
and get

\be
\Im I_{ij}=-\frac{2}{3}\frac{\delta_{ij}}{(2\pi)^2}\sum_n^\infty(e^{n\beta|q|}E_1(n\beta|q|)-e^{-n\beta|q|}E^*(n\beta|q|)),
\ee
where $E_1(x)=\int_x^\infty dt e^{-t}/t $ and $E^*(x)=-P\int_{-x}^\infty dt e^{-t}/t \;$ ($P$ stands for the principal value).
Now we use the following property of the above functions \cite{erdelyi},

\be
\int_0^\infty\frac{t\cos(xt)}{a^2+t^2}\,dt=\frac{1}{2}[e^{ax}E_1(ax)-e^{-ax}E^*(ax)] \, ,
\ee
and get

\be
\sum_n^\infty(e^{n\beta|q|}E_1(n\beta|q|)-e^{-n\beta|q|}E^*(n\beta|q|))=2\int_0^\infty\,dt t \cos(\beta|q|t)\sum_n\frac{1}{n^2+t^2}
\, .
\ee
The sum can be carried out using complex variable techniques. We obtain
\be
\Im I_{ij}=\frac{2}{3}\frac{\delta_{ij}q}{(2\pi)^2}\int_0^\infty\frac{\cos(\beta|q|t)\,dt}{t\tanh(\pi t)}(\tanh(\pi t)-\pi t)
\ee
Finally, the integral 
yields,

\be
\Im I_{ij}=\frac{2}{3}\frac{\delta_{ij}q}{(2\pi)^2}(\ln(\frac{2\pi}{\beta|q|})+\Re\psi(\frac{i\beta|q|}{2\pi})),
\ee
where $\psi(x)=\Gamma'(x)/\Gamma (x)$

\subsection{Computation of (\ref{vl})}

We derive here the result displayed in (\ref{vl}). A similar
computation has been done in the past for QCD \cite{Voloshin:1979uv}. However, (\ref{vl}) cannot be obtained by just taking the Abelian limit  of the QCD result. The latter is singular because it does not contain the projector ${\bar P}_n$. We will proceed in a way analogous to \cite{Voloshin:1979uv}, but keeping $E\not=E_n$ in the terms in which the limit $E\rightarrow E_n$ does not exist.
We set $E=E_n+\lambda$ with $\lambda\to 0$. If we drop 
$\bar{P_n}$ in the numerator of (\ref{vl}) we get

\be
\langle n|r^i\frac{1}{H-E}r_i|n\rangle=\frac{f(n,l)}{\lambda}+g(n,l)+O(\lambda) .
\ee
If we included ${\bar P}_n$ in the numerator we get an exact cancellation of $f(n,l)$, so that the limit $\lambda\to 0$ can be taken safely:

\be
\langle n|r^i\frac{1}{H-E}r_i|n\rangle=\int\,d^3x\,d^3y\langle n|r^i|x\rangle\langle x|\frac{1}{H-E}|y\rangle\langle y|r_i|n\rangle ,
\ee
we use $r^i|x\rangle=x^i|x\rangle$ and 
the following formula \cite{Voloshin:1979uv}

\be
\langle x|\frac{1}{H-E}|y\rangle=\sum_{l=0}^\infty(2l+1)G_l(x,y;E)P_l(\frac{x^iy_i}{xy}),
\ee
with

\be
G_l(x,y;-k^2/(2m))=\frac{mk}{\pi}(2kx)^l(2ky)^le^{-k(x+y)}\sum_{s=0}^\infty\frac{L_s^{2l+1}(2kx)L_s^{2l+1}(2ky)s!}{(s+l+1-Zm\alpha/k)(s+2l+1)!}.
\label{total}
\ee
Note that if $E=E_n$ then $Zm\alpha/k=n$, so we will have a pole at $s=n-l-1$. For the angular integration 
we take into account that $x^iy_i=xy(\frac{x^iy_i}{xy})$ and combine it with $P_l$ using

\be
xP_l(x)=\frac{(l+1)P_{l+1}(x)+lP_{l-1}(x)}{2l+1} .
\ee
We get finally

\bea
\langle n|r^i\frac{1}{H-E}r_i|n\rangle =\frac{2mkl(n-l-1)!}{n(2k_0)^5(2l+1)(n+l)!}\sum_{s=0}^\infty\frac{s!(I(l-1,s))^2}{(s+l-Zm\alpha/k)(s+2l-1)!}+\\
\frac{2mk(l+1)(n-l-1)!}{n(2k_0)^5(2l+1)(n+l)!}\sum_0^\infty\frac{(s!(I(l+1,s))^2}{(s+l+2-Zm\alpha/k)(s+2l+3)!}, \nonumber
\eea
with

\be
I(h,s)=2k_0\int\,dx(2k_0x)^{l+3}(2kx)^he^{-(k+k_0)x}L^{2l+1}_{n-l-1}(2k_0x)L^{2h+1}_s(2kx).
\label{i}
\ee
We  define $k_0$ so that, $E_n=-k_0^2/2m$. For terms in the sum that are well defined when $k\rightarrow k_0$
we can just put $k=k_0$. For the terms in the sum that are singular 
we have to expand for small $\lambda$ and then subtract the singular part as we mentioned before. This is indeed what the introduction of ${\bar P}_n=1-P_n$ does.
In order to demonstrate it 
let us look at the $P_n$ part.

\be
\langle n|r^i\frac{P_n}{H-E}r_i|n\rangle=-\frac{1}{\lambda}\int\,d^3x\,d^3y\langle n|r^i|x\rangle\langle x|P_n|y\rangle\langle y|r_i|n\rangle .
\ee
We will proceed in an analogous way as in the calculation with no projector above, so that 
the cancellation will become apparent:

\be
\langle x|P_n|y\rangle=\sum_{l=0}^{n-1}(2l+1)G_l(x,y)P_l(\frac{x^iy_i}{xy}),
\ee
where

\be
G_l(x,y)=\frac{mk_0}{\pi}(2k_0x)^l(2k_0y)^le^{-k_0(x+y)}\frac{k_0^2}{mn}\frac{(n-l-1)!L_{n-l-1}^{2l+1}(2k_0x)L_{n-l-1}^{2l+1}(2k_0y)}{(n+l)!},
\label{pole}
\ee
is similar to $G_l(x,y;-k^2/2m)$ above ,
but the summation for $s$ is restricted to singular terms. 
By comparing (\ref{pole}) and (\ref{total}) we can easily see that the $1/\lambda$ terms cancel even before doing the radial integration. Since the $P_n$ part is proportional to $1/\lambda$ (no finite pieces), we only have to calculate the finite contribution (in an expansion in $\lambda$) to the part with no projector. This can be easily obtained by expanding $k$ about $k_0$ in (\ref{i}) (recall that the derivative of a Laguerre polynomial is a Laguerre polynomial).
The computation can be terminated in an analytic form using,

\be
\int_0^\infty\,dxe^{-x}L_n^k(x)L_{n'}^{k'}(x)x^s=s!\sum_{r=0}^{{\rm min}[n,n']}(-1)^{n+n'+r}{{s-k}\choose{n-r}}{{s-k'}\choose{n'-r}}{{-s-1}\choose r}.
\ee

\section{Integrals in terms of special functions}

We give here the definitions of various functions appearing  
in the paper in terms of one parameter integrals and provide expressions in terms of special functions.
The Debye mass 
can be expressed as

\be
m_D^2:=\frac{8m^2}{(2\pi)^2}e^2(2f(m\beta)+h(m\beta)),
\ee
where

\be
f(m\beta):=\frac{1}{m^2}\int\,dk\frac{k^2}{\sqrt{k^2+m^2}(e^{\beta\sqrt{k^2+m^2}}+1)}=-\sum_{n=1}^\infty(-1)^n\frac{K_1(n\beta m)}{n\beta m},
\ee

\be
h(m\b ):=\int_0^\infty dk {1\over\sqrt{k^2+m^2}\left( e^{\b\sqrt{k^2+m^2}}+1\right)}=
-\sum_{n=1}^\infty(-1)^nK_0(n\beta m),
\ee
and $g(m\b )$ 
as

\be
g(m\b ):=
\b^2\int_0^\infty dk {k\over e^{\b\sqrt{k^2+m^2}}+1}=m\beta\ln(1+e^{m\beta})+Li_{2}(-e^{m\beta})+{\pi^2\over 6}-{m^2\beta^2\over 2},
\ee
$K_0(x)$ and $K_1(x)$ are Bessel functions and $Li_{2} (x)$ the dilogarithmic function.

\section{Matching the vertex function}

\subsection{Matching QED to NRQED+HTL}

In order to carry out the matching for the vertex function ($\G$) we have to deal with the doubling of degrees of freedom.
There are three external particles in the vertex, and each one can be of type 1 or type 2 (following the notation of \cite{lebellac}), so $\Gamma$ is a tensor with eight components. But, because of kinematic constraints it cannot happen that there is and internal photon on shell and an internal electron on shell at the same time, so the only components that are nonvanishing are $111$, $121$, $212$ and $222$ (the middle index corresponds to the photon). If we take into account that the matrix elements of the propagators in the real-time formalism are not independent,
we obtain $\Gamma_{111}=\Gamma_{222}$ and $\Gamma_{121}=\Gamma_{212}$. Notice also that, for the physics of an atom with an infinitely heavy nucleus, the only components that have a contribution at first order are $111$ and $212$.

As we did with the self-energy, we calculate first the contribution from 
the thermal photons. In this case
$\Gamma_{212}=0$, so we only have to calculate the 111 component. The calculation is done 
by matching three-point Green functions in QED and NRQED:

\bea
\parbox{30mm}{\fmfreuse{tree}}+\parbox{30mm}{\fmfreuse{vertex}}+\parbox{30mm}{
\begin{fmffile}{./article1}
\begin{fmfgraph}(80,80)
\fmfleft{i}
\fmfright{o}
\fmfbottom{b}
\fmf{fermion}{i,v}
\fmf{dashes}{v,b}
\fmf{fermion}{v,o1}
\fmf{fermion}{o1,o2}
\fmf{photon,left,tension=0.1}{o1,o2}
\fmf{fermion}{o2,o}
\end{fmfgraph}
\end{fmffile}}+\parbox{30mm}{
\begin{fmffile}{./article2}
\begin{fmfgraph}(80,80)
\fmfleft{i}
\fmfright{o}
\fmfbottom{b}
\fmf{fermion}{i,i1}
\fmf{fermion}{i1,i2}
\fmf{photon,left,tension=0.1}{i1,i2}
\fmf{fermion}{i2,v}
\fmf{dashes}{v,b}
\fmf{fermion}{v,o}
\end{fmfgraph}
\end{fmffile}} = \label{vqed}\\ 
Z_\psi[\parbox{30mm}{\fmfreuse{tree}}+\parbox{30mm}{
\begin{fmffile}{./article3}
\begin{fmfgraph*}(80,80)
\fmfleft{i}
\fmfright{o}
\fmfbottom{b}
\fmf{fermion}{i,v}
\fmfv{label=$c_D$
,l.a=90
}{v}
\fmf{fermion}{v,o}
\fmf{dashes}{v,b}
\end{fmfgraph*}
\end{fmffile}}
+\parbox{30mm}{
\begin{fmffile}{./article4}
\begin{fmfgraph*}(80,80)
\fmfleft{i}
\fmfright{o}
\fmfbottom{b}
\fmf{fermion}{i,v}
\fmfv{label=$c_S$
,l.a=90
}{v}
\fmf{fermion}{v,o}
\fmf{dashes}{v,b}
\end{fmfgraph*}
\end{fmffile}}].
\eea
The first row represents QED diagrams (all of them are sandwiched between the projectors $\frac{1+\gamma_0}{2}$), and the second one represents NRQED diagrams. We find

\be
\delta e=0,
\ee
\be
c_D=1+\frac{8\alpha}{3\pi}(\ln\frac{\beta m}{2\pi}+\gamma-\ln 2+\frac{5}{6}+O(T^2)),
\ee
\be
\delta c_S=0+O(\alpha T^2).
\ee
The finite-temperature contribution to $c_F$ is irrelevant at first order for a hydrogen atom 
since the corresponding operator contains tranverse photons only. We have restricted ourselves to calculate the leading order contribution to $c_D$ and $c_S$ in the limit of $T\ll m$, which is enough for illustration purposes. This is also justified because for $T\sim m$ the neglected contributions, as well as the ones taken into account, produce modifications of the spectrum of order $m\alpha^5$, whereas we will see in the following that there are  contributions from
the
thermal fluctuations of the electrons 
at order $m\alpha^2$, which will modify the physics of the hydrogen atom in a much more profound way.

We focus next in the contributions from the thermal electrons. The leading order contribution in QED comes from 
the second diagram in (\ref{vqed})

\be
\delta \Gamma_{111}=i8\pi^2\alpha me\int_0^1\frac{\,dw}{(1-w^2)^{3/2}(e^{\frac{\beta m}{\sqrt{1-w^2}}}+1)}\int\frac{\,d\Omega}{(2\pi)^3}(1-\frac{1}{e^{\frac{\beta m}{\sqrt{1-w^2}}}+1})\delta(q_0-{\bf q}{\bf w}) , \label{G111}
\ee

\be
\delta \Gamma_{212}=i4\pi^2\alpha me\int_0^1\frac{\,dw}{(1-w^2)^{3/2}(e^{\frac{\beta m}{\sqrt{1-w^2}}}+1)}\int\frac{\,d\Omega}{(2\pi)^3}(1-\frac{1}{e^{\frac{\beta m}{\sqrt{1-w^2}}}+1})\delta(q_0-{\bf q}{\bf w}) ,
\ee
($w=k/\sqrt{k^2+m^2}$, $k$ being the momentum circulating in the loop). Note that this contribution is nonlocal and cannot be matched to any of the NRQED operators. It can be matched to the following nonlocal operator

\be
\delta 
L=\int\,d^3{\bf w} f(w) {\bf w}{\bf E}\frac{1}{{\bf w}{\bf \nabla}}\delta(i\partial_0-i{\bf w}{\bfnabla})\psi^+\psi,
\label{vnl}
\ee
where

\be
f(w)=\frac{\alpha em}{\pi^2}\frac{1}{e^{\frac{\beta m}{\sqrt{1-w^2}}}+1}\frac{1}{w^2}\frac{1}{(1-w^2)^{3/2}}\left(1-\frac{1}{e^{\frac{\beta m}{\sqrt{1-w^2}}}+1}\right).
\ee
(${\bf E}$ is the electric field). This operator becomes as important as the leading order Lagrangian when $q\sim m\alpha$, and it is suppressed only by $e$ when $q\sim me$, the scale of the Debye mass. Hence the thermal fluctuations of the electrons have a bigger impact in the NRQED Lagrangian than any of the relativistic or radiative corrections.

\subsection{Matching to pNRQED and cancellation of the scale dependence}

The nonlocal vertex above can easily be matched to pNRQED at tree level by expanding the energy over the three-momentum. At leading order we have

\be
\delta 
L_{pNRQED}=\int\,d^3{\bf w} f(w) {\bf w}{\bf E}\frac{1}{{\bf w}{\bf \nabla}}\delta(-i{\bf w}{\bfnabla})\psi^+\psi.
\ee 
This vertex is IR divergent, so in the calculations in pNRQED there should appear an ultraviolet divergence, in order to get a cancellation of the $\mu$ dependency. It indeed appears in a diagram of the type,

\be
\parbox{30mm}{
\begin{fmffile}{./article5}
\begin{fmfgraph}(80,80)
\fmfleft{i}
\fmfright{o}
\fmfbottom{b}
\fmf{fermion}{i,v1}
\fmf{fermion}{v1,v2}
\fmf{dashes}{v2,b}
\fmf{fermion}{v2,v3}
\fmf{dashes,left,tension=0.1}{v1,v3}
\fmf{fermion}{v3,o}
\end{fmfgraph}
\end{fmffile}
} , \label{usuv}
\ee
where the internal lines are now nonrelativistic propagators for the electrons and HTL propagators for the photons\footnote{Note that Coulomb resummations can be ignored at the scale of the Debye mass ($me$), since they only become important for momentum transfer of the order $m\a$ or smaller.}.

For simplicity, let us check this cancellation in a specific piece of the tensor vertex (for the remaining pieces it will be analogous). We focus on $\d \G_{111}$, in the case $q_0\to 0$, and take into account only the temperature-dependent part in one of the electron propagators and the zero-temperature part in the other one, which will be enough for illustration purposes. Let us call it $\Gamma^*$.

From the NRQED matching we have (from the first term in (\ref{G111}), by taking $q_0=0$ and undoing the change of variable $w=k/\sqrt{k^2+m^2}$)

\be
\Gamma^*=-\frac{e^3\pi}{(2\pi)^2|{\bf q}|}\int_0^\infty\frac{\,dk\sqrt{k^2+m^2}}{k(e^{\beta\sqrt{k^2+m^2}}+1)}.
\ee
Since we are interested in only the IR divergent behavior, we may substitute the integrand by the following regulated expression

\be
\Gamma^*\sim-\frac{e^3\pi m}{(2\pi)^2|{\bf q}|(e^{\beta m}+1)}\mu^{-\epsilon}\int_0^\infty\frac{\,dk e^{-k/m}}{k^{1-\epsilon}}=-\frac{e^3\pi}{(2\pi)^2|{\bf q}|}\frac{m}{e^{\beta m}+1}(\frac{1}{\epsilon}+\ln(\frac{m}{\mu})).
\label{muNRQED}
\ee
Any calculation in pNRQED+HTL involving the contribution above will also involve the diagram (\ref{usuv}) with nonrelativistic propagators for the electrons and HTL propagators for the ($A_0$) photons. Let us take into account only the temperature-dependent part in one of the two electron propagators and the zero temperature part in the other one, in accordance with the evaluation of $\G^\ast$ above, and call the corresponding contribution ${\tilde \Gamma}^*$. We have,

\be
{\tilde\Gamma}^*=\int\frac{\,d^4 k}{(2\pi)^4}(-ie)\frac{(-2\pi)\delta(k_0-\frac{k^2}{2m})}{e^{\beta|m+k_0\vert}+1}(-ie)\frac{i}{q_0+k_0-\frac{({\bf q}+{\bf k})^2}{2m}+i\eta}(-ie)
\D_{11}\left(\vert {\bf p}-{\bf k}\vert , p_0-k_0\right).
\ee
Due to the $\delta$ function and to the fact that $p_0\sim {\bf p}^2/2m\ll p$, we can use the expression (\ref{D11}) for $\D_{11}$.
We focus on the UV behavior of the expression above, since we are only interested in identifying the $\mu$ dependence, which should cancel that of (\ref{muNRQED}). We can then neglect the imaginary part of $\D_{11}$, which leads to finite expressions, and approximate

\begin{eqnarray}
{\tilde\Gamma}^*&\sim& e^3m\int\frac{\,d^3{\bf k}}{(2\pi)^3}\frac{1}{e^{\beta m}+1}\frac{1}{-({\bf q}{\bf k})+i\eta}
\frac{i}{({\bf k})^2+m_D^2}\nn\\
&=&
\frac{-\pi e^3 m}{(e^{\beta m}+1)|{\bf q}|}\frac{1}{(2\pi)^2}\int_0^\infty\frac{\,dk k^{1+\epsilon}}{
k^2
+m_D^2}\\
&=&\frac{-\pi e^3 m}{(e^{\beta m}+1)|{\bf q}|}\frac{1}{(2\pi)^2}\left(-\frac{1}{\epsilon}+\frac{1}{2}\ln\left(\frac{\mu^2}{
m_D^2
}\right)\right)\nn .
\end{eqnarray}
In the second relation we have carried out the angular integration and introduced DR (neglecting $\e$ in the finite pieces). 
If we add ${\Gamma}^*$ to ${\tilde\Gamma}^*$ we see that the $\mu$ dependence indeed cancels, as it should.


\begin{thebibliography}{99}
\bibitem{farley}
J. W. Farley and W. H. Wing
Phys.\ Rev.\ A {\bf 23}, 2397 (1981). 
\bibitem{PalanquesMestre:1984xw}
A.~Palanques-Mestre and R.~Tarrach,
Phys.\ Rev.\  D {\bf 30}, 502 (1984).
\bibitem{Donoghue:1984zz}
J.~F.~Donoghue, B.~R.~Holstein and R.~W.~Robinett,
Annals Phys.\  {\bf 164}, 233 (1985)
[Erratum-ibid.\  {\bf 172}, 483 (1986)].
\bibitem{Cha:1984tz}
B.~Y.~Cha and J.~H.~Yee,
Phys.\ Rev.\  D {\bf 32}, 1038 (1985).
\bibitem{hollberg}
L.~Hollberg and J.~L.~Hall,
Phys. Rev. Lett. {\bf 53}, 230 (1984).
\bibitem{Brambilla:2004wf}
CERN report No.2005-005, 2005 Quarkonium Working Group, N.~Brambilla {\it et al.}
\bibitem{Caswell:1985ui}
W.~E.~Caswell and G.~P.~Lepage,
Phys.\ Lett.\  B {\bf 167}, 437 (1986).
\bibitem{Pineda:1997bj}
A.~Pineda and J.~Soto,
Nucl.\ Phys.\ Proc.\ Suppl.\  {\bf 64}, 428 (1998).
\bibitem{Pineda:1997ie}
A.~Pineda and J.~Soto,
Phys.\ Lett.\  B {\bf 420}, 391 (1998).
\bibitem{Pineda:1998kn}
A.~Pineda and J.~Soto,
Phys.\ Rev.\  D {\bf 59}, 016005 (1999).
\bibitem{Czarnecki:1999mw}
A.~Czarnecki, K.~Melnikov and A.~Yelkhovsky,
Phys.\ Rev.\  A {\bf 59}, 4316 (1999).
\bibitem{Manohar:2000rz}
A.~V.~Manohar and I.~W.~Stewart,
Phys.\ Rev.\ Lett.\  {\bf 85}, 2248 (2000).
\bibitem{Pineda:2002bv}
A.~Pineda,
Phys.\ Rev.\  A {\bf 66}, 062108 (2002).
\bibitem{Brambilla:2004jw}
N.~Brambilla, A.~Pineda, J.~Soto and A.~Vairo,
Rev.\ Mod.\ Phys.\  {\bf 77}, 1423 (2005).
\bibitem{Matsui:1986dk}
T.~Matsui and H.~Satz,
Phys.\ Lett.\  B {\bf 178}, 416 (1986).
\bibitem{Lourenco:2006sr}
C.~Lourenco,
Nucl.\ Phys.\  A {\bf 783}, 451 (2007).
\bibitem{Laine:2006ns}
M.~Laine, O.~Philipsen, P.~Romatschke and M.~Tassler,
JHEP {\bf 0703}, 054 (2007).
\bibitem{Laine:2007gj}
M.~Laine,
JHEP {\bf 0705}, 028 (2007).
\bibitem{Burnier:2007qm}
Y.~Burnier, M.~Laine and M.~Vepsalainen,
JHEP {\bf 0801}, 043 (2008).
\bibitem{Beraudo:2007ky}
A.~Beraudo, J.~P.~Blaizot and C.~Ratti,
Nucl.\ Phys.\  A {\bf 806}, 312 (2008)
\bibitem{na}
N. Brambilla, J. Ghiglieri, A. Vairo, P. Petreczky, 
Phys.\ Rev.\  D {\bf 78}, 014017 (2008)
\bibitem{lebellac}
M. Le Bellac,
``Thermal Field Theory,''
{\it  Cambridge, UK: Cambridge University Press (1996) 256 p.}
\bibitem{Manohar}
A.~V.~Manohar,
Phys.\ Rev.\  D {\bf 56} (1997) 230.
\bibitem{Frenkel:1989br}
J.~Frenkel and J.~C.~Taylor,
Nucl.\ Phys.\  B {\bf 334}, 199 (1990).
\bibitem{Braaten:1990az}
E.~Braaten and R.~D.~Pisarski,
Nucl.\ Phys.\  B {\bf 339}, 310 (1990).
\bibitem{Taylor:1990ia}
J.~C.~Taylor and S.~M.~H.~Wong,
Nucl.\ Phys.\  B {\bf 346}, 115 (1990).
\bibitem{Braaten:1991gm}
E.~Braaten and R.~D.~Pisarski,
Phys.\ Rev.\  D {\bf 45}, 1827 (1992).
\bibitem{Voloshin:1979uv}
M.~B.~Voloshin,
Sov.\ J.\ Nucl.\ Phys.\  {\bf 36}, 143 (1982)
[Yad.\ Fiz.\  {\bf 36}, 247 (1982)].
\bibitem{Pineda:1998id}
A.~Pineda, Ph.D. Thesis, Universitat de Barcelona, 1998
\bibitem{erdelyi}
A.~Erderlyi
``Higher transcendental functions''
{\it McGraw-Hill, 1953-1955}
\bibitem{Pisarski:1997cp}
R.~D.~Pisarski,
e-print arXiv:hep-ph/9710370.
\bibitem{Thoma:2000dc}
M.~H.~Thoma,
e-print arXiv:hep-ph/0010164.
\bibitem{Titard:1993nn}
S.~Titard and F.~J.~Yndurain,
Phys.\ Rev.\  D {\bf 49}, 6007 (1994).
\bibitem{GarciaiTormo:2005bs}
X.~Garcia i Tormo and J.~Soto,
Phys.\ Rev.\ Lett.\  {\bf 96}, 111801 (2006).
\end{thebibliography}
\end{document}